\documentclass[letterpaper,conference,10pt]{IEEEtran}
\IEEEoverridecommandlockouts
\usepackage{color}
\usepackage[utf8]{inputenc}
\usepackage{amsmath}
\usepackage{amssymb}
\usepackage{amsfonts}
\usepackage{mathtools}
\usepackage{booktabs}
\usepackage{algorithmic}
\usepackage{array}
\usepackage[font=footnotesize,caption=false]{subfig}
\usepackage{url}
\usepackage{cite}
\usepackage{color}
\usepackage{graphicx}
\usepackage{bm}
\usepackage{chngcntr}
\counterwithout{paragraph}{subsubsection}

\newtheorem{conjecture}{Conjecture}

\newtheorem{theorem}{Theorem}

\newcommand\fixme[1]{{\em\bf{[FIXME: #1]}}}

\newcommand{\FF}{\mathbb{F}}
\newcommand\bv[1]{{\bf #1}}
\newcommand\co[1]{}

\makeatletter
\def\ps@headings{%
\def\@oddhead{\mbox{}\scriptsize\rightmark \hfil \thepage}%
\def\@evenhead{\scriptsize\thepage \hfil \leftmark\mbox{}}%
\def\@oddfoot{}%
\def\@evenfoot{}}
\makeatother
\usepackage{geometry}
\begin{document}

\title{RapidRAID: Pipelined Erasure Codes for \\Fast Data Archival in Distributed Storage Systems\thanks{L. Pamies-Juarez and F. Oggier's research is supported by the Singapore National Research
Foundation under Research Grant NRF-CRP2-2007-03. A. Datta's work is supported by A*Star TSRP grant number 1021580038.}}
\author{\IEEEauthorblockN{Lluis Pamies-Juarez, Anwitaman Datta and Frederique Oggier}
\IEEEauthorblockA{Nanyang Technological University (Singapore)\\
Email: \{lpjuarez,anwitaman,frederique\}@ntu.edu.sg}}

\maketitle
\begin{abstract}
To achieve reliability in distributed storage systems, data
has usually been replicated across different nodes. However the increasing
volume of data to be stored has motivated the introduction of erasure codes,
a storage efficient alternative to replication, particularly suited for
archival in data centers, where old datasets (rarely accessed) can be erasure
encoded, while replicas are maintained only for the latest data.  Many recent
works consider the design of new storage-centric erasure codes for improved
repairability. In contrast, this paper addresses the \emph{migration} from
replication to encoding: traditionally erasure coding is an \emph{atomic operation}
in that a single node with the whole object encodes and uploads all the
encoded pieces. Although large datasets can be concurrently archived by
distributing individual object encodings among different nodes, the network
and computing capacity of individual nodes constrain the archival process due
to such atomicity.

We propose a new \emph{pipelined coding strategy} that distributes the network and
computing load of single-object encodings among different nodes, which also
speeds up multiple object archival. We further present \emph{RapidRAID codes}, an
explicit family of pipelined erasure codes which provides fast archival
without compromising either data reliability or storage overheads. Finally,
we provide a real implementation of RapidRAID codes and benchmark its
performance using both a cluster of 50 nodes and a set of Amazon EC2
instances. Experiments show that RapidRAID codes reduce a single
object's coding time by up to 90\%, while when multiple objects are encoded
concurrently, the reduction is up to 20\%.
\end{abstract}

\section{Introduction}

Networked distributed storage systems such as Google file-system
(GFS)~\cite{googlefs}, Amazon S3~\cite{s3} or Hadoop file-system
(HDFS)~\cite{hdfs} spread data among several storage nodes and allow to scale
out from hundreds to thousands of commodity storage servers able to
accommodate the ever-growing volume of data to be stored. To ensure that data
survives failures of some of the storage nodes, all data needs to be
redundantly stored. The simplest way to introduce redundancy is to store
multiple copies (or replicas) of each data across the system. But erasure
codes, a more sophisticated type of redundancy, can provide equivalent or even
better fault-tolerance than replication for significantly lower storage
overhead~\cite{highavail}, and hence have increasingly been embraced in 
recent times in systems such as Microsoft Azure \cite{azureec}, Hadoop
FS~\cite{warehouseface,diskreduce} and the new version of the Google File
System \cite{ecgoogle} among others. Typical choices of erasure codes used
in these systems have an overall overhead of $1.3\times$--$1.5\times$ the
size of the original data~\cite{azureec,hdfsraid,diskreduce}, which allows to
reduce up to 50\% the typical overhead of storing three replicas.

Although erasure codes have the potential to significantly reduce storage
costs in distributed storage systems, there are still two advantages of using
replication to store newly introduced data:

\begin{itemize}
\item {\em Pipelined Insertion:} Replication allows to easily pipeline the
redundancy generation process: data being stored in a node can be
simultaneously forwarded to a second node, and from this second node to a
third, and so on~\cite{hdfs,googlefs}. Such pipelining process allows to
distribute the redundancy generation costs among different nodes, achieving a
high storage throughout as well as an immediate data reliability.

\item {\em Data Locality:} Freshly introduced data in the system is very
likely to be accessed and used, e.g., in a batch process to carry out some
analytics.  Replicating the data in several storage nodes allows the task
scheduler to exploit data-locality: jobs are scheduled on the same nodes
where data is located \cite{mapreduce,facebook}. Such a scheduling strategy
reduces network latencies and increases data and processing throughputs.

\end{itemize}

Due to these properties, distributed storage systems often store newly
introduced data using replication, and rely on erasure codes to archive older
and infrequently accessed data~\cite{warehouseface,diskreduce,diskreduce2}.
Such a pragmatic design allows systems to enjoy the benefits of replication
(fast data insertion, data locality, etc.) when the data is in frequent use,
as well as that of erasure codes (high fault-tolerance for lower storage
cost) when the data is not accessed regularly, but still needs
to be preserved.

The need to access a specific stored data reduces significantly within a
short period of time \cite{diskreduce,diskreduce2}, which justifies replacing
the replicas by an erasure code based archival. This migration usually
consists of an atomic operation where a single storage node obtains the
entire data object (by downloading blocks from different nodes if needed),
encodes it, and finally uploads various parity blocks to different storage
nodes~\cite{diskreduce}, after which the number of replicas can be safely
reduced to one. Although the encoding of one data object using this naive
approach is inherently centralized, large datasets (containing several data
objects) can sometimes be concurrently encoded by distributing individual
encoding operations across different nodes. This does not change the fact
that the limited network and computing capacity of individual nodes remain a
bottleneck that slows down the whole archival process.

While different aspects of erasure coding based distributed storage systems
have been studied recently, which include maximizing the fault-tolerance of
erasure codes~\cite{datasurvival,hetawarej}, reducing the costs of repairing
failures~\cite{Li,OD,kermarrec}, or deduplicating encoded
data~\cite{dedupec}, this paper instead looks at a relatively unexplored
problem, that of the efficiency of the migration from replication to erasure
codes, aiming at optimizing the data archival in distributed storage systems.

Our main contributions are three-fold.

(1) We propose a novel coding strategy that splits the single-object encoding
operation into different tasks that can be concurrently executed in different
nodes, thus distributing the network and computing load of the archival
process across multiple nodes, which in turn speeds up the archival process.
Our new encoding scheme is inspired by the \emph{pipelined insertions} used
in replication:
First, the encoding process is distributed among those nodes storing
replicated data of the object to be encoded, which exploits \emph{data locality} and
saves network traffic. We then arrange the encoding nodes in a
pipeline where each node sends some partially encoded data to the next node,
which creates parity data simultaneously on different storage nodes,
avoiding the extra time required to distribute the parity after the encoding
process is terminated.

(2) We further present RapidRAID codes, an explicit family of erasure codes
that realizes the pipelined erasure coding idea and provides fast archival
without compromising either on data reliability or on storage overhead.
Interestingly, RapidRAID codes only require the existence of two object
replicas to execute a pipeline encoding, which makes them suitable for
archiving data in reduced redundancy systems. Additionally, RapidRAID codes
offer flexible parameter choices to realize different storage overheads (up
to 2$\times$ the size of the original data) and different data reliability
guarantees. \co{ In specific, the RapidRAID framework allows to construct
codes with a maximum storage overhead of 2$\times$ the size of the original
data, and obtains maximum data reliabilities for storage overheads of the
order of $1.3\times$ the size of the original object.  However, higher
storage overheads can be achieved by slightly relaxing data reliability, but
still achieving a data reliability guarantee higher than replication and
comparable to classical erasure codes. For example, we show how for a
practical code configuration~\cite{azureec}, i.e., a (16,11) code with a
storage overhead of $1.45\times$, RapidRAID codes achieve better data
reliability than replication, and the same than classical erasure codes.}

(3) We finally provide a real implementation of RapidRAID codes that we
benchmark both in a small cluster of 50 HP ThinClients as well as in a set of
Amazon EC2 virtual instances. Our experimental results show that RapidRAID
coding reduces the coding time of single data objects by up to 90\%, and by
up to 20\% for batch processing the coding of multiple objects. The benefits
of RapidRAID codes are also visible when part of the network is congested.
The presence of congested nodes has less detrimental effects on RapidRAID
encoding times than on traditional encoding times.

The rest of the paper is organized as follows. In Section~\ref{s:background}
we provide the basic background on distributed storage systems and classical
erasure codes. In Section~\ref{s:times} we estimate the coding times of
classical erasure codes and show how pipelined erasure coding speeds up the
coding time by exploiting data locality. In Sections~\ref{s:rrex}
and~\ref{s:rrge} we present the family of RapidRAID codes and we
experimentally evaluate its performance in Section~\ref{s:eval}.  Finally,
Sections~\ref{s:relwork} and~\ref{s:concl} respectively present the related
work and our conclusions.

\section{Background on Erasure Codes} \label{s:background}

Distributed storage systems used in data centers have started to adopt a
hybrid strategy for redundancy, where replicas of the newly inserted data are
created, while erasure codes are preferred for archival of the same data once
it does not need to be regularly accessed anymore, but still needs to be
preserved. The number of replicas is then reduced. The use of erasure coding
for archival increases the fault tolerance of the system while reducing
storage overheads with respect to replication~\cite{azureec}, though
replication remains so far the best form of redundancy for new data since it
is likely to be frequently manipulated.

Formally, the encoding process takes $k$ blocks of data and computes $m$
parity blocks (or redundancy blocks), which will be stored in $m$ other
different storage nodes. In most cases, since it is unlikely to find data
objects that were exactly split into $k$ blocks during the insertion process,
the $k$ blocks used in the encoding process might belong to different data
objects. For example, in some systems files from the same directory are
jointly encoded~\cite{diskreduce}.

An optimal erasure code in terms of the trade-off between storage overhead
and fault tolerance is called a maximum distance separable (MDS) code, and
has the property that the original object can be reconstructed from any $k$
out of the $n=k+m$ stored blocks, tolerating the loss of any $m=n-k$ blocks.
The notation ``$(n,k)$ code'' is often used to emphasize the code parameters.
Examples of the most widely used MDS codes are the Reed-Solomon codes. Such
codes will be referred to as {\em classical erasure codes}, to distinguish
them from newly designed erasure codes.

We will denote a data object to be stored by a vector $\bv o=(o_1,\dots,o_k)$
of $k\times l$ bits, that is each $o_i$, $i=1,\ldots,k$, is a string of $l$
bits. Operations are typically performed using finite field arithmetic, that
is, the two bits $\{0,1\}$ are seen as forming the finite field
$\mathbb{F}_2$ of two elements, while $o_i$, $i=1,\ldots,k$, then belong to
the binary extension field $\mathbb F_{2^l}$ containing $2^l$ elements.
Encoding of the object $\bv o$ is performed using an ($n\times k$) generator matrix $G$
such that $G\cdot{\bv o}^T = {\bv c}^T$, to obtain an $n$-dimensional codeword
${\bv c}=(c_1,\dots,c_n)$, of size $n\times l$ bits.  When the generator
matrix $G$ has the form $G=[I_k,G']^T$ where $I_k$ is the identity matrix and
$G'$ is a $k\times (n-k)$ matrix, the codeword ${\bv c}$ becomes ${\bv
c}=[{\bv o},{\bv r}]$ where $\bv o$ is the original object, and ${\bv r}$
contains the $m\times l$ bits of redundancy. The code is then said to be
systematic, in which case the $k$ parts of the original object remain
unaltered after the coding process. The data can then still be read without
requiring a decoding process.

Due to the computational complexity of finite field arithmetic, erasure codes
usually need to operate on small fields (with small $l$ values) to guarantee
a fast coding process. Usually $\FF_{2^8}$ or $\FF_{2^{16}}$ ($l=8$ or
$l=16$) are preferred due to their efficient
manipulation using 8-bit and 16-bit CPU words. However, the size of the field
also constrains the size of the object (which is of $lk$ bits) to be either
$8k$ or $16k$ bits long, for relatively small values of $k$. In distributed
storage systems where the $k$ blocks are usually tens of megabytes long, the
coding is handled per part. The coding process iteratively takes $k$ input
words ($l$ bit words) from each of the $k$ original blocks to form a small
object of size $lk$, which can be easily encoded.

\section{Pipelining the Redundancy Generation Process} \label{s:times}

One of the main drawbacks of classical erasure codes is that the encoding
process is an atomic operation in that a single node has the responsibility to
download $k$ blocks (from any of the existing replicas), encode them, and
finally upload the resulting $m$ parity blocks to $m$ other nodes
\cite{diskreduce}. In this case the encoding node becomes a network and
computing bottleneck that slows down the whole coding process.

\begin{figure}
\centering
\includegraphics[scale=0.9]{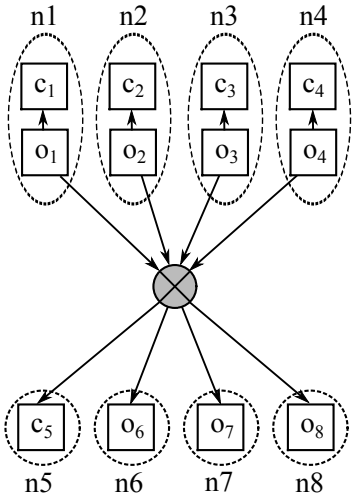} \caption{Network flow
required to encode a data object using a classical systematic (8,4) erasure code
using nodes n1$\dots$n8. The $\otimes$ symbol denotes a coding operation.}
\label{f:pec}
\end{figure}

To understand better why such an atomicity results in long
encoding times, we depict in Fig.~\ref{f:pec} an example of an object
encoding using a classical systematic ($8,4$) erasure code: an object ${\bv
o}=(o_1,o_2,o_3,o_4)$ is encoded into an $8$-dimensional codeword \linebreak
${\bv c}=(c_1,\ldots,c_8)=({\bv o},c_5,\ldots,c_8)$. The node $i$ (denoted by n$i$ on
the figure) stores a replica of the raw data block $c_i=o_i$, $i=1,\ldots,4$. To migrate to an
erasure encoded data, the node executing the encoding process (denoted by
$\otimes$) downloads the $k=4$ original blocks from any of the
existing replicas (here from node $1,\ldots,4$), and computes the
redundancy blocks $c_5,\ldots,c_8$ which are then uploaded to nodes $5$ to
$8$. The number of transmitted blocks is $n=8$, and it could have been
reduced to $n-1=7$ if the coding process were run for example in node $4$,
which already stored $c_4$ locally. In this toy example, exploiting data
locality could save a block transmission.

To analytically obtain an estimate of the time required for encoding one
object using a classical erasure code, we consider the best possible scenario
and assume that the coding process is done in a streamlined manner, meaning
that the coding node downloads in parallel all the $k$ original blocks and
starts to generate parity data immediately after receiving the first few
bytes from each of the $k$ source nodes (e.g., once the first $k$ network
buffers are filled).  Concurrently with the encoding of this data, the coding
node continues to receive data from the $k$ source nodes, and uploads the
partially generated parity data to the $m-1$ destination nodes. The time
required to encode an object can then be approximated by:

\begin{equation} T_\text{classical} = \tau_\text{block}\cdot\max\{k,m-1\} +
\tau_\text{classical}, \label{e:tclassic} \end{equation}

\noindent where $\tau_\text{block}$ is the time needed to download a single
data block under normal network conditions, and $\tau_\text{classical}$
represents the time required to generate parity data from the first $k$
network buffers.  Since the size of the blocks to be encoded are relatively
large we will assume that the time required to transfer a block between two
nodes is several orders of magnitude longer than the time required to
partially encode an amount of data equivalent to the size of a network
buffer: i.e., $\tau_\text{block} \gg \tau_\text{classical}.$

\begin{figure}
\centering
\includegraphics[scale=0.9]{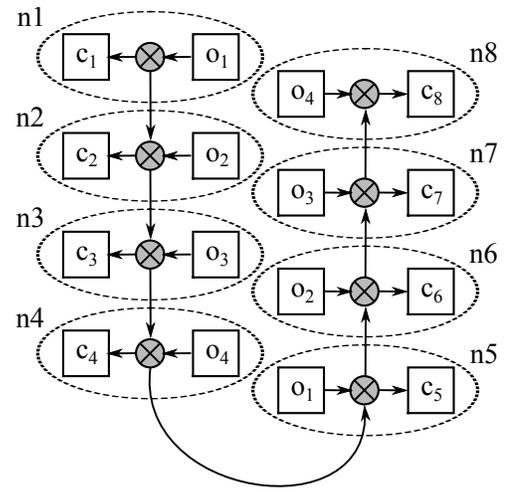} \caption{Network flow
required to encode a data object using a (8,4) pipelined erasure code using
node n1$\dots$n8. The $\otimes$ symbol denotes a coding operation.}
\label{f:ppl}
\end{figure}

One way to avoid the bottleneck of having a single coding node is to pipeline
the creation of erasure code redundancy and distribute the redundancy
generation costs among different storage nodes. The main idea behind our
pipelined strategy is to take advantage of the fact that the data to be
encoded is already spread and replicated over different nodes. Then, each of
the nodes with one of the replicas can combine the data it stores with data
from other nodes to generate part of the final codeword $\bv c$. In
Fig.~\ref{f:ppl} we depict an example of this idea using the same code
parameters (8,4) used in Fig.~\ref{f:pec}, though here we do not insist on
the code being systematic. Nodes $1$ to $4$ store together a replica of
the stored object $\bv o$ as before (that is node $i$ stores $o_i$,
$i=1,\ldots,4$) but this time nodes $5$ to $8$ store a second replica of
the same object as well. The coding process proceeds as follows. The first
node sends a multiple of $o_1$ to the second node. The node $2$ computes a linear
combination of this multiple of $o_1$ with $o_2$ and forwards the result to the node $3$.
The node $3$ has now its own data $o_3$, and again computes a linear
combination of $o_3$ with what it received. The process is iteratively
repeated from node $i$ to node $i+1$, $i=1,\ldots,7$. Simultaneously to
this pipeline process, each node also generates its own redundancy block
$c_i$, based on what it owns and receives, which does not have to be the same
linear combination as that sent to the next node. The set of all the locally
generated blocks constitutes the final codeword $\bv c=(c_1,\dots,c_8)$.

Assuming that only two replicas of $\bv o$ are used in the process, the
maximum length of the final codeword should be constrained to $n=2k$,
although we will see in next sections that any $n\leq2k$ is possible.
Additionally, note that the coding process only requires to transmit seven
temporal blocks (in general $n-1$ blocks are transmitted) which entails the
same network traffic as a classical encoding process. However, the coding
time for the pipelined strategy is significantly reduced. In this case we can
measure the coding time as the time required to transmit one block
$\tau_\text{block}$, plus $n-1$ times the delay taken to receive and
encode a network buffer, denoted by $\tau_\text{pipe}$ (we assume here the
same streamlined coding strategy):

\begin{equation}
T_\text{pipe} = \tau_\text{block} + (n-1)\tau_\text{pipe}.
\label{e:tpipe}
\end{equation}

\noindent Similarly, due to the large size of the blocks being encoded (of the
order of tens of megabytes) we can also assume that the time required to
transfer a block between two nodes is several orders of magnitude longer than
the time required to partially encode an amount of data equivalent to the
size of a network buffer: $\tau_\text{block} \gg \tau_\text{pipe}$. However,
since $\tau_\text{block}\gg\tau_\text{pipe}$ and $\tau_\text{block} \gg
\tau_\text{classical}$, it is easy to see when we compare (\ref{e:tclassic})
and (\ref{e:tpipe}) that the factor $\max\{k,m-1\}$ in (\ref{e:tclassic})
makes $T_\text{classical}$ several times larger than $T_\text{pipe}$. In
Section~\ref{s:eval} we will support this claim with real experiments.

One possible criticism of the pipelined coding strategy is that unlike in
classical erasure codes, the generated codeword does not contain a raw copy
of the original data (i.e., it is not a systematic code). The immediate
consequence is that accessing stored data will always require a decoding
operation, which always comes with an associated CPU overhead.  However, the
benefits of a fast and less CPU-demanding encoding process (as we will see in
Section VI) outweighs the relative inefficiency of data access, since the
latter is infrequent.  Furthermore, empirical studies have shown how erasure
encoded data can be accessed with relatively low latencies, even when data
needs to be decoded~\cite{azureec}, and this latency can be further
ameliorated by adopting pipelined decoding operations (faster than classical
decoding operations), which are not reported here because of space
restrictions.

\section{RapidRAID: Motivating Examples} \label{s:rrex}

In this section we present RapidRAID codes, an explicit family of erasure
codes that realize the idea of pipelined erasure codes presented in the
previous section. We first illustrate the code construction through two simple
examples, and in Section~\ref{s:rrge} we formalize the definition of
RapidRAID codes.

\subsection{Example for $n=2k$}

We continue with an $(8,4)$ erasure code, as used in the previous section. An
object ${\bv o}=(o_1,o_2,o_3,o_4),~o_i\in\FF_{2^l}$, of $k=4$ blocks is
stored over $n=8$ nodes using a codeword ${\bf c}=(c_1,\ldots,c_8)$, and
two replicas of ${\bv o}$ are initially scattered as follows
(this is the same original placement as that of Fig.~\ref{f:ppl}):
\[
\begin{array}{llll}
\mbox{node 1:}~o_1,&\mbox{ node 2:}~o_2,&\mbox{ node 3:}~o_3,&\mbox{ node 4:}~o_4, \\
\mbox{node 5:}~o_1,&\mbox{ node 6:}~o_2,&\mbox{ node 7:}~o_3,&\mbox{ node 8:}~o_4.
\end{array}
\]
Based on this replica placement, we split the RapidRAID coding process in two phases:

\vspace{1mm}
\noindent {\bf Phase 1 (vertical coding)}: Following the pipeline depicted in
Fig.~\ref{f:ppl}, node $1$ forwards some multiple of $o_1$ to node 2, which
computes a linear combination of the received data with $o_2$, and forwards
it again to node 3, and so on. More generally, node $i$ encodes
the data it gets from the previous node together with the data it already has
and forwards it to the next node. We denote the data forwarded from node $i$
to its successor, node $j$, by $x_{i,j}$, which is defined as follows:
\begin{align*}
x_{1,2} &=~ o_1\psi_1, \\
x_{2,3} &=~ x_{1,2} + o_2\psi_2 = o_1\psi_1 + o_2\psi_2, \\
x_{3,4} &=~ x_{2,3} + o_3\psi_3 = o_1\psi_1 + o_2\psi_2 + o_3\psi_3, \\
x_{4,5} &=~ x_{3,4} + o_4\psi_4 \\&=~ o_1\psi_1 + o_2\psi_2 + o_3\psi_3 + o_4\psi_4, \\
x_{5,6} &=~ x_{4,5} + o_1\psi_5 \\&=~ o_1(\psi_1+\psi_5) + o_2\psi_2 + o_3\psi_3 + o_4\psi_4, \\
x_{6,7} &=~ x_{5,6} + o_2\psi_6 \\&=~ o_1(\psi_1+\psi_5) + o_2(\psi_2+\psi_6) + o_3\psi_3 + o_4\psi_4, \\
x_{7,8} &=~ x_{6,7} + o_3\psi_7 \\&=~ o_1(\psi_1+\psi_5) + o_2(\psi_2+\psi_6) + o_3(\psi_3+\psi_7) + o_4\psi_4,
\end{align*}
where $\psi_j\in\FF_{2^l}$, $j=1,\ldots,7$, are predetermined values.

\vspace{1mm}
\noindent {\bf Phase 2 (horizontal coding)}:
Each of the $n$ involved nodes also generates an element of the final
codeword $c_i$ by encoding the received data together with the locally stored
data as follows:
\begin{align*}
c_1 =&~ o_1\xi_1, \\
c_2 =&~ x_{1,2} + o_2\xi_2 = o_1\psi_1 + o_2\xi_2, \\
c_3 =&~ x_{2,3} + o_3\xi_3 = o_1\psi_1 + o_2\psi_2 + o_3\xi_3, \\
c_4 =&~ x_{3,4} + o_4\xi_4 = o_1\psi_1 + o_2\psi_2 + o_3\psi_3 + o_4\xi_4, \\
c_5 =&~ x_{4,5} + o_1\xi_5 \\=&~ o_1(\psi_1+\xi_5) + o_2\psi_2 + o_3\psi_3 + o_4\psi_4, \\
c_6 =&~ x_{5,6} + o_2\xi_6 \\=&~ o_1(\psi_1+\psi_5) + o_2(\psi_2+\xi_6) + o_3\psi_3 + o_4\psi_4, \\
c_7 =&~ x_{6,7} + o_3\xi_7 \\=&~ o_1(\psi_1+\psi_5) + o_2(\psi_2+\psi_6) + o_3(\psi_3+\xi_7) + o_4\psi_4, \\
c_8 =&~ x_{7,8} + o_4\xi_8 \\=&~ o_1(\psi_1+\psi_5) + o_2(\psi_2+\psi_6) + o_3(\psi_3+\psi_7)+o_4(\psi_4+\xi_8),
\end{align*}
where $\xi_j\in\FF_{2^l}$, $j=1,\ldots,8$, are also predetermined values.

Although we defined the coding process using two logically different phases, we want to
highlight that when the coding process is implemented as a streamlined
process, both phases can be executed simultaneously: as soon as node $i$
receives the first few bytes of $x_{i-1,i}$ it can start generating the first
bytes of $c_i$, and concurrently forward $x_{i,i+1}$ to node $i+1$.

\subsection{Object Reconstruction and Fault Tolerance}
\label{s:ftol}

Using the notation of Section~\ref{s:background}, we can express
the RapidRAID coding process of the (8,4) example using the standard linear
coding notation $G\cdot{\bv o}^T = {\bv c}^T$ as
$$
\begin{bmatrix*}[l]
\xi_1 & 0 & 0 & 0 \\
\psi_1 & \xi_2 & 0 & 0 \\
\psi_1 & \psi_2 & \xi_3 & 0 \\
\psi_1 & \psi_2 & \psi_3 & \xi_4 \\
\psi_1+\xi_5 & \psi_2 & \psi_3 & \psi_4 \\
\psi_1+\psi_5 & \psi_2+\xi_6 & \psi_3 & \psi_4 \\
\psi_1+\psi_5 & \psi_2+\psi_6 & \psi_3+\xi_7 & \psi_4 \\
\psi_1+\psi_5 & \psi_2+\psi_6 & \psi_3+\psi_7& \psi_4+\xi_8
\end{bmatrix*} \cdot
\begin{bmatrix} o_1 \\ o_2 \\ o_3 \\ o_4
\end{bmatrix} =
\begin{bmatrix} c_1 \\ c_2 \\ c_3 \\ c_4 \\ c_5 \\ c_6 \\ c_7 \\ c_8
\end{bmatrix}.
$$
It is easy to see that we can use the Gauss elimination
method to reconstruct the original object, $\bv o$, from any subset of four
linearly independent symbols of $\bv c$.  Maximizing the number of linearly
independent 4-subsets in $\bv c$ can be done by exhaustive computational
search of the values taken by $\psi_i$ and $\xi_i$ once the size $2^l$ of the
field is fixed. When all $k$-subsets in $\bv c$ are linearly independent, the
code then becomes MDS, which achieves the highest possible fault tolerance
given any $k$ and $n$.  Note that the larger the field $\FF_{2^l}$, the more
likely it is to remove the linear dependencies within $\bv
c$~\cite{randomstorage}.

However, even by selecting the optimal values of $\psi_i$ and $\xi_i$, there
could be some intrinsic dependencies introduced by the pipelined coding
process itself that cannot be removed. In the example of the
(8,4) code proposed, from all the $\binom{8}{4}=70$ possible 4-subsets, there
is one single linearly dependent $4$-subset, namely $\{c_1,c_2,c_5,c_6\}$,
which cannot be removed, no matter the values taken by $\xi_i$ and $\psi_i$
in $\FF_{2^l}$, for any $l$.  Recall that $2 \equiv 0$ in $\FF_{2^l}$.  Then
the following linear combination of $c_1,c_2,c_5$ and $c_6$ always evaluates to zero:
\begin{eqnarray*}
\!\!&\!\!&\!\!c_1\left[(\psi_1\xi_6\xi_2^{-1}+\psi_5+\xi_5)\xi_1^{-1}\right] + c_2\left[\xi_6\xi_2^{-1}\right] + c_5 + c_6 \\
\!\!&\!\!=&\!\!o_1\xi_1(\psi_1\xi_6\xi_2^{-2}+\psi_5+\xi_5)\xi_1^{-1} + (o_1\psi_1+o_2\xi_2)\xi_6\xi_2^{-1} \\
\!\!&\!\!&\!\!+(o_1\psi_1+o_1\xi_5+o_2\psi_2+o_3\psi_3+o_4\psi_4) \\
\!\!&\!\!&\!\!+(o_1\psi_1+o_1\psi_5 +o_2\psi_2+o_2\xi_6+o_3\psi_3+o_4\psi_4)\\
\!\!&\!\!=&\!\!o_1\xi_1\psi_1\xi_6\xi_2^{-1}\xi_1^{-1}\!+o_1\xi_1\psi_5\xi_1^{-1}+o_1\xi_1\xi_5 \xi_1^{-1}\!+ o_1\psi_1\xi_6\xi_2^{-1}\\
\!\!&\!\!&\!\!+o_2\xi_2\xi_6\xi_2^{-1}+o_1\psi_1+o_1\xi_5+o_2\psi_2+o_3\psi_3+o_4\psi_4 \\
\!\!&\!\!&\!\!+ o_1\psi_1+o_1\psi_5 +o_2\psi_2+o_2\xi_6+o_3\psi_3+o_4\psi_4 = 0.
\end{eqnarray*}
It shows that the code is not an MDS code: if all the redundant blocks but
$\{c_1,c_2,c_5,c_6\}$ fail, it will be impossible to recover the original
data $\bv o$. In Section~\ref{s:rrge} we will analyze in detail which are the
($n,k$) values that allow to obtain MDS codes, and for the rest of ($n,k$)
values, we will quantify the impact that the non-MDS property has on
the overall data reliability.

\subsection{Example for $n<2k$}

The previous (8,4) code example enjoys a symmetric construction
inherited from $n=2k$, but we can extend the RapidRAID coding scheme for $n\leq 2k$. As an example we consider the case of a (6,4)
code, which requires replicas of $\bv o$ to be
initially overlapped on the $n=6$ nodes as follows:
\[
\begin{array}{lll}
\mbox{node 1:}~o_1,&\mbox{ node 3:}~o_3,o_1,&\mbox{ node 5:}~o_3,\\
\mbox{node 2:}~o_2,&\mbox{ node 4:}~o_4,o_2,&\mbox{ node 6:}~o_4,\\
\end{array}
\]
The rest of the coding process continues as previously explained. The
basic difference will be on the computation made by nodes 3 and 4, which in
this case corresponds to:
\begin{eqnarray*}
&x_{3,4} = x_{2,3} + o_3\psi_3 + o_1\psi_4, ~~&c_3 = x_{2,3} + o_3\xi_3 + o_1\xi_4,\\
&x_{4,5} = x_{3,4} + o_4\psi_5 + o_2\psi_6, ~~&c_4 = x_{3,4} + o_4\xi_5 + o_1\xi_6.
\end{eqnarray*}
Note that some of the subindexes of coefficients $\psi$ and $\xi$ might need
to be altered accordingly.

\section{RapidRAID: General Definition} \label{s:rrge}

Inspired by the examples of previous section, we now present a general
definition of RapidRAID codes for any pair ($n,k$) of parameters, where
$n\leq2k$. We start by stating the requirements that RapidRAID imposes on how
data must be stored:
\begin{itemize}
\item As shown in the (6,4) example code, when $k<2k$ two of the stored
replicas should be overlapped between $n$ storage nodes: a replica of
$\bv o$ should be placed in nodes $1$ to $k$, and a second replica of $\bv o$
in nodes from $n-k$ to $n$.

\item The final $n$ redundancy blocks forming $\bv c$ have to be generated
(and finally stored) in nodes that were already storing a replica of the
original data.
\end{itemize}

We then formally define the temporal redundant block that each node $i$
in the pipelined chain sends to its successor as:
\begin{equation}
x_{i,i+1} = x_{i-1,i} + \sum_{o_j\in\rm{node}~i}o_j\psi_i, ~1<i<n-1,
\label{e:pfwd}
\end{equation}
with $x_{0,1}=0$, while the final redundant block $c_i$ generated/stored in
each node $i$ is:
\begin{equation}
c_i = x_{i-1,i} + \sum_{o_j\in\rm{node}~i}o_j\xi_i, ~1<i<n,
\label{e:pcode}
\end{equation}
where $\psi_i,\xi_i\in\FF_{2^l}$ are static predetermined values
specifically chosen to guarantee maximum fault tolerance.

\subsection{Fault Tolerance Analysis}

As we already mentioned, the fault tolerance of the code depends on the
number of linearly independent blocks within the codeword $\bv c$. Optimally,
if the code is MDS, all the $\binom{n}{k}$ $k$-subsets of $\bv c$ are linearly independent.
In practice, achieving the MDS
property is not always possible due to different types of linear dependencies
generated during the construction of the RapidRAID code. We distinguish two
different types of these linear dependencies:
\begin{enumerate}
\item {\bf Natural dependencies} are introduced by the pipelined coding process
itself and cannot be removed, no matter the values taken by
$\xi_i$ and $\psi_i$.
\item {\bf Accidental dependencies} appear due to a
bad choice of the values of $\xi_i$ and $\psi_i$.
\end{enumerate}

\begin{figure}
\centering
\subfloat[Percentage of linearly independent $k$-subsets.]%
{\includegraphics{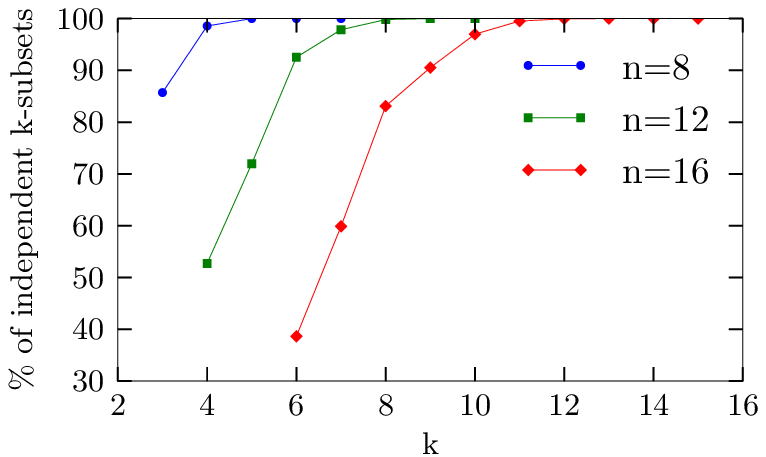}\label{f:perc}}
\vspace{2mm}
\quad
\subfloat[Number of linearly dependent $k$-subsets.]%
{\includegraphics{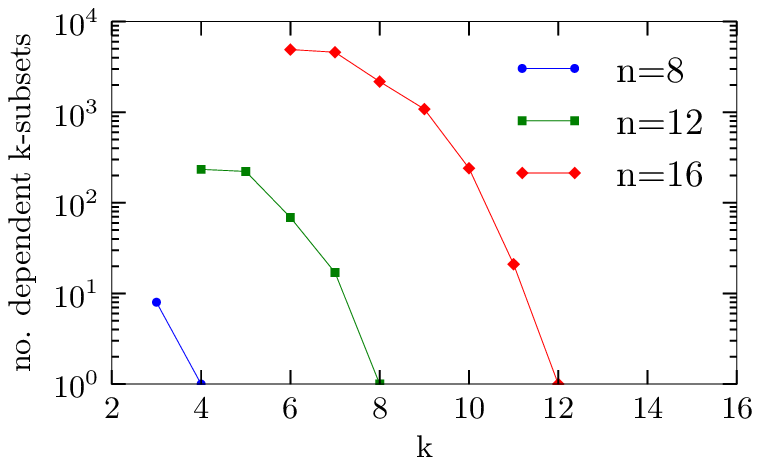}\label{f:num}}
\caption{Evaluation of the linear dependencies in ($n,k$)
RapidRAID codewords. We consider three $n$ values with all the possible $k$
values such that $\frac{n}{2}\leq k<n$.}
\label{f:numeric}
\vspace{-3mm}
\end{figure}

To evaluate the fault tolerance of an $(n,k)$ RapidRAID code, we need to
count the different linear dependencies in its codewords. We first
analytically detect \emph{natural dependencies} by enumerating all the
possible $k$-subsets, and for each $k$-subset, we determine by symbolic
computation whether it contains linear dependencies. Once we know that there
is no linear dependency, we pick values of $\psi_i$ and $\xi_i$ so as to
avoid {\em accidental dependencies}. This can be done at random for
relatively large fields such as $\FF_{2^{16}}$, where almost any random set
of coefficients guarantees the absence of accidental
dependencies~\cite{randomstorage}.  For small fields like $\FF_{2^{8}}$,
finding a set of coefficients without accidental dependencies might require
long exhaustive searches.

Such an enumeration of all possible $k$-subsets is feasible only for small
values of $n$, due to the fast growth of the number $\binom{n}{k}$ of
$k$-subsets to test.  In Fig.~\ref{f:numeric}, we computed the number of
natural linear dependencies of $(n,k)$ RapidRAID codes with
$n\in\{8,12,16\}$, and all the possible values of $k$, $\frac{n}{2}\leq k<n$.
In Fig.~\ref{f:perc} we show the percentage of linearly independent
$k$-subsets and in Fig.~\ref{f:num} the absolute number of linearly dependent
$k$-subsets. We observe that RapidRAID codes achieve the MDS property when
$k\geq n-3$.

After analyzing all the RapidRAID codes for $n\leq16$, we propose the
following conjecture:
\vspace{1mm}
\begin{conjecture}
An $(n,k)$ RapidRAID code as defined by (\ref{e:pfwd}) and (\ref{e:pcode}) is
maximum distance separable (MDS) if $k\geq n-3$.
\end{conjecture}

\vspace{1mm}\noindent However, we would like to highlight that some of the
non-MDS codes (when $k<n-3$) still achieve high percentages of linearly
independent $k$-subsets. This is the case for example of a (16,11) RapidRAID
code, evaluated later in this paper.

To complete the fault tolerance analysis of $(n,k)$ RapidRAID codes, we now
consider their static resilience, which is the probability of being able to
reconstruct a given stored object when a fraction $p$ of random storage nodes
fail.  This static resilience for different node failure probabilities using
the ``number of 9's'' metric\footnote{For example, `three nines' represents a
probability of 0.999.} is shown in Table~\ref{t:reliab}, where we compare
three different codes: (i) a (16,11) RapidRAID code, which is non-MDS, (ii) a
(16,11) classical MDS code, and (iii) the standard replication scheme with
three replicas. We see that although the static resilience of the RapidRAID
code is slightly lower than the classical erasure code, for storage systems
with low node failure probabilities ($p\leq0.01$), RapidRAID codes achieve at
least the same resiliency as the de-facto standard 3-way replication scheme.
According to data center studies published in~\cite{Schroeder,Pinheiro}, the
annualized failure rate (AFR) of modern hard disk drives (HDD) is in the
range of 2\% to 5\%, depending on the age of the disk. Since the time
required to repair a disk failure (which includes the time to detect the disk
failure plus the time to repair and restore the missing data) is in the range
of minutes or a few hours~\cite{Schroeder}, it is reasonable to expect less
than 1\% of simultaneous disk failures, making the RapidRAID codes family an
attractive alternative to replace classical erasure codes in data centers.
Besides, the actual trend in datacenters is to use solid state disks (SSD),
which have even lower AFRs as compared to traditional HDD. Further note that
the actual chance of data loss is much lower than the values indicated by
static resilience analysis if the system is repaired and thus faults are not
allowed to accumulate.

\begin{table}
\centering
\caption{Static resiliency of three different redundancy schemes (in number
of 9's) for different probabilities of node failure $p$.}
\label{t:reliab}
\begin{tabular}{lcccc} \toprule
 & \boldmath$p\!\!=\!\!0.2$ & \boldmath$p\!\!=\!\!0.1$ &
 \boldmath$p\!\!=\!\!0.01$ & \boldmath$p\!\!=\!\!0.001$ \\ \toprule
3-replica system & 2 & 3 & 6 & 9 \\ \midrule
(16,11) classical EC & 1 & 2 & 8 & 14 \\ \midrule
(16,11) RapidDAID & 0 & 2 & 6 & 11 \\ \bottomrule
\end{tabular}
\vspace{-6mm}
\end{table}

\section{Evaluation}
\label{s:eval}

In this section we evaluate the coding performance of a RapidRAID code and
compare it with that of classical erasure coding. The code that we choose for
the evaluation is a (16,11) code, with parameters similar to those used in
real distributed storage systems~\cite{azureec}, which offer a data
reliability comparable to a (16,11) classical erasure code (see Table~\ref{t:reliab}).

\subsection{Implementation and Testbed}

In order to fairly compare coding times of RapidRAID codes with those of
classical erasure codes, we developed an experimental distributed storage
system\footnote{Available online:
\url{https://github.com/llpamies/ClusterDFS}} which consists of a fast Python
server infrastructure providing basic store/retrieve operations, as well as
finite field arithmetic required to encode and forward data in pipelined
erasure codes. The finite field arithmetic is implemented using the
Jerasure~\cite{jerasure} library, which contains a fast set of functions
(optimized C code) designed to construct efficient erasure codes.

Over our distributed storage system we integrated two different erasure
codes:

\begin{itemize}
\item A (16,11) classical Reed-Solomon erasure code using Cauchy generator
matrices, as it is already implemented in the Jerasure library.  We adjust
the erasure code parameters to guarantee maximum performance as it is
suggested in~\cite{Plank09}, which makes the Cauchy Reed Solomon code to
clearly outperform other open source erasure coding libraries~\cite{Plank09}.
We will refer to this code implementation as \emph{CEC} (\emph{Classical
Erasure Code}).

\item A (16,11) RapidRAID code implemented using the finite field arithmetic
from Jerasure. This implementation can either work with 8 bit or 16 bit
arithmetic, with operations in $\FF_{2^8}$ or $\FF_{2^{16}}$ respectively. In
each case the values of all $\phi_i$ and $\xi_i$ coefficients are chosen to
maximize the obtained fault tolerance. We will refer to the 8bits and 16bits
RapidRAID implementations as \emph{RR8} and \emph{RR16} respectively.

\end{itemize}

In the case of \emph{RR8} the use of a small finite field makes it
very difficult to find coefficient values guaranteeing the absence
of accidental linear dependencies. In this case, the 8bit (16,11) RapidRAID
implementation achieves data reliability values slightly lower than the ones
depicted in Table~\ref{t:reliab}. Despite this lower reliability, we
include the 8bit implementation in our evaluation to show the effects that
the word size has in coding times. Note that our RapidRAID
implementation also includes a fast pipelined decoding mechanism that is
not discussed here because of space restrictions.

We evaluate the three coding settings, \emph{CEC}, \emph{RR8} and \emph{RR16}
in both a small cluster of 50 HP t5745 ThinClient computers, and a set of 16
small instances in the Amazon EC2 cloud computing service. We will refer to
the ThinClient and Amazon EC2 testbed as {\em TPC} and {\em EC2}
respectively. Finally, in all the experiments we assume that the size of all
the $k=11$ original blocks is of 64MB, which is the default block size in GFS
and HDFS~\cite{googlefs,hdfs}. It means that the size of the original object
to be stored is of 704MB (11$\times$64MB), and the final erasure encoded
object takes 1024MB (16$\times$64MB), which represents a storage overhead of
approximately $1.45\times$ the size of the original data.

\subsection{Computing Resource Usage}

\begin{table}
\centering
\caption{Overall coding time of three (16,11) code implementations.}
\label{t:cpus}
\begin{tabular}{m{4cm}ccc} \toprule
{\bf CPU} & {\bf CEC} & {\bf RR8} & {\bf RR16} \\ \toprule
Intel Atom (N280) \newline 1.66GHz; 512KB cache &
17.81 &
5.06 &
27.33 \\ \midrule
Intel Xeon (E5645) \newline 2.40GHz; 12,288KB cache &
5.20 &
3.50 &
4.31 \\ \midrule
Intel Core2 Quad (Q9400) \newline 2.66GHz; 3,072KB cache &
4.13 &
1.47 &
1.95 \\ \bottomrule
\end{tabular}
\vspace{-6mm}
\end{table}

Before evaluating coding times we will measure the overall computing
requirements of the three evaluated codes. This metric is of special interest
in datacenters where an archiving process requiring little overall computing
resources is preferred due to the low interference it has on the normal
datacenter operations.

To measure the overall computing requirements of the {\em CEC} implementation
we execute an encoding process where the $k=11$ original blocks and the $m=5$
parity are all stored in the local file system, avoiding all the network I/O.
In that case the encoding time corresponds basically to the time the CPU is
dedicated to execute the coding operations. Similarly, to measure the overall
computing requirements of the {\em RR8} and {\em RR16} implementations, we
run an encoding process where the execution of the $n\!=\!16$ nodes occur in
a single node, avoiding also all the network I/O.

In Table~\ref{t:cpus} we depict the average encoding time of the three
encoding implementations when all the computing is executed in a single node
and no network communication is involved. We show the results for three
different CPUs. The first case (Intel Atom) corresponds to the execution time
in the Thinclient computers, the second case (Intel Xeon) is an Amazon EC2
\emph{small instance}, and the last one (Intel Core2) a personal desktop
computer. Except in the case of Atom, both RapidRAID implementations require
less CPU time to encode the same amount of data (i.e., 704MB) than the {\em
CEC} implementation. In the case of the Atom CPU, due to the small
size of the cache memory, the Jerasure library cannot allocate the whole
lookup table required to perform $\FF_{2^{16}}$ arithmetic, which increases
{\em RR16} coding times as compared to {\em RR8}.

We observe that RapidRAID codes can be computed faster than even one of the
fastest implementation of classical erasure codes, and thus its impact on CPU
usage is favorable.

\subsection{General Coding Times}

\begin{figure}
\centering
\subfloat[Time required to encode a single object using a (16,11) code.]%
{\includegraphics{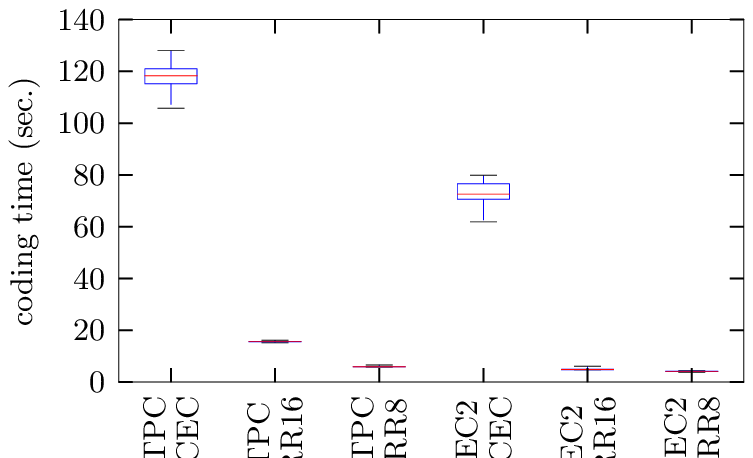}\label{f:single}}
\vspace{3mm}
\quad
\subfloat[Time required to concurrently encode 16 objects using a (16,11) code.]%
{\includegraphics{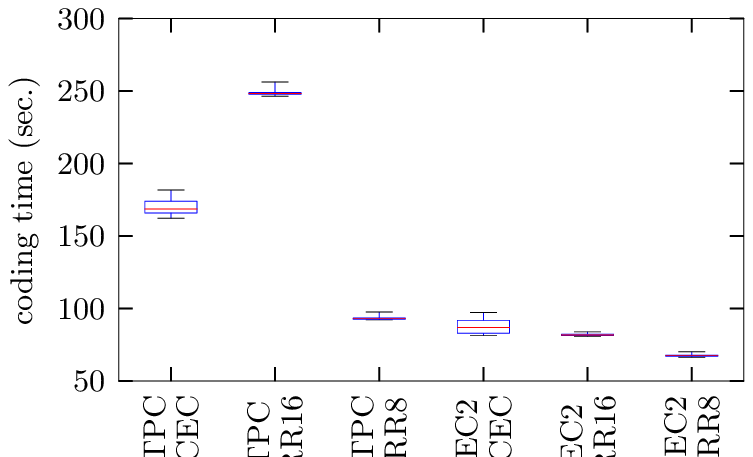}\label{f:parallel}}
\caption{Coding times of the three different code implementations. Each
candle depicts the median value, the 25-75\% percentiles and the max-min
values.}
\label{f:generaltimes}
\vspace{-2mm}
\end{figure}

In Fig.~\ref{f:generaltimes} we measure the encoding times of the three
different implementations for a single object encoding, as well as multiple
object encodings.

In Fig.~\ref{f:single} a single data object is encoded in a totally idle
system. We see how the two RapidRAID implementations have of the order of
90\% shorter coding times as compared to the classical erasure code
implementation. In this case, by distributing the network and computing load
of the encoding process across 16 different nodes, RapidRAID codes
significantly speed up the single data object's archival process.

However, this speedup is obtained at the expense of involving 16 nodes in the
encoding process. It is then interesting to measure the encoding throughput
of a classical erasure code involving the same number of nodes, i.e., when 16
encoding process are executed in parallel. In Fig.~\ref{f:parallel} we depict
the per-object encoding times obtained by executing 16 concurrent classical
encoding processes and 16 RapidRAID encoding processes on a group of 16
nodes. In the {\em EC2} setting, the two RapidRAID implementations achieve a
reduction of the overall coding time by up to 20\%. On the Thinclients, the
16bit RapidRAID implementation requires around 50\% longer coding times than
classical erasure codes due to problems with the small cache size.

\subsection{Coding Times in Congested Networks}

In practice, storage nodes might be executing other tasks concurrently with
data archival processes, which might cause some nodes to experience network
congestions that in turn might affect the coding times. Although nodes in the {\em
EC2} setting are already virtual computers subjected to real network
congestions, we needed to be able to arbitrary reduce the network capacity of
some nodes to evaluate the potential effects that severe network congestions
can have on RapidRAID coding times.
To evaluate such effects of congestion, we use
the Linux \emph{netem} driver to introduce arbitrary congestions in our
cluster of ThinClients.
Specifically, we use {\em netem} to reduce the network bandwidth of some
nodes from 1GBps to 500MBps, and add to these nodes a 100ms network latency
(with a deviation of up to $\pm 10$ms).

\begin{figure}
\centering
\subfloat[Encoding a single object using a (16,11) code.]%
{\includegraphics{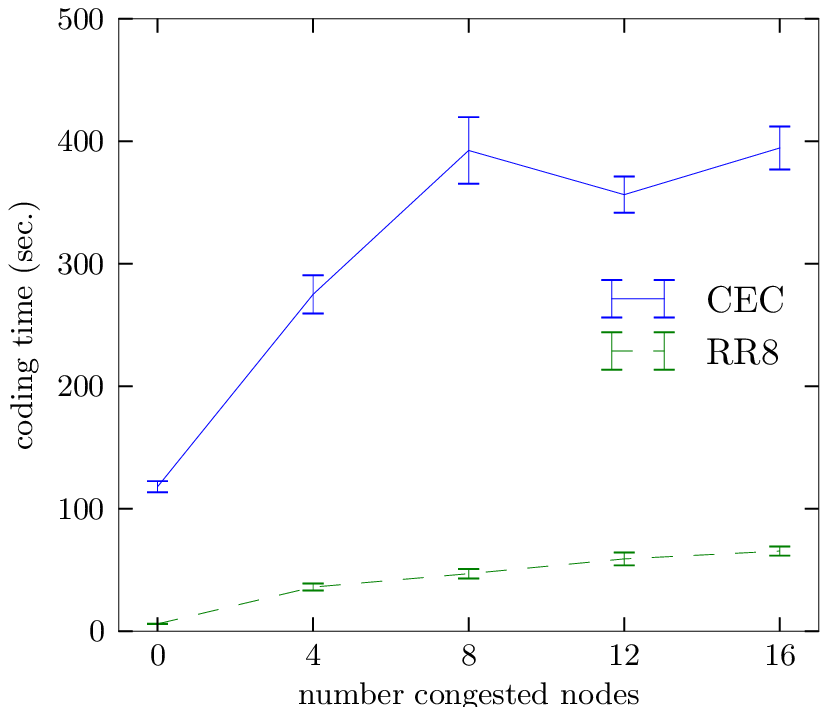}\label{f:congsingle}}
\vspace{3mm}
\quad
\subfloat[Concurrently encode 16 objects using a (16,11) code.]%
{\includegraphics{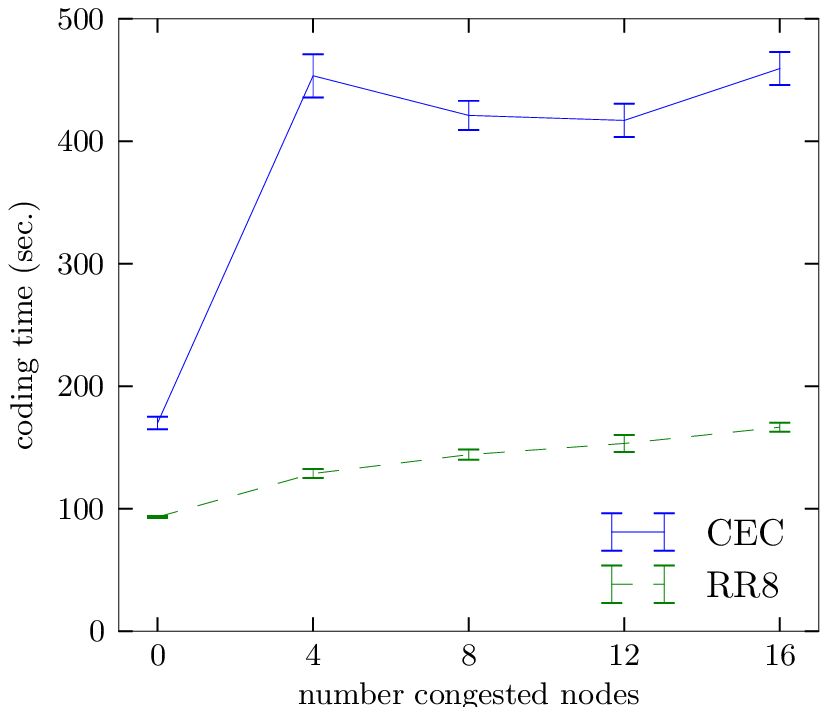}\label{f:congparallel}}
\caption{Average time required to encode a 16 concurrent objects using a
(16,11) Cauchy Reed-Solomon code and a 8bit (16,11) RapidRaid code. Nodes
have 500Mbps connections with a latency of 100ms$\pm$10ms. Error bars depict
the standard deviation value.}
\label{f:cong}
\vspace{-2mm}
\end{figure}

In Fig.~\ref{f:cong} we depict the effects that different network congestion
levels have in coding times of the {\em CEC} and {\em RR8} implementations.
Note that we only use the 8bits RapidRAID implementation due to the
impossibility to run efficient $\FF_{2^{16}}$ arithmetic in the ThinClient
cluster. In Fig.~\ref{f:congsingle} we show the time required to
encode a single object. In the case of RapidRAID codes, coding times have a
quasi-linear behavior when the number of congested nodes increases.  However,
in the case of classical erasure codes, we can see how a single congested
node has major impacts to the coding times.  Similarly, in
Fig.~\ref{f:congparallel} we depict the per-object coding times of 16
concurrently encoded objects. Compared with the single object coding time,
the presence of a single congested node has even more impact on the coding
times of classical erasure codes.  In general these results show how
classical erasure codes have a worse resilience to congested networks than
RapidRAID codes.

\section{Related Work}
\label{s:relwork}
Despite widespread use of erasure coding for archiving data in distributed
storage systems, existing literature does not explore the process of
migration from replication based redundancy to erasure code based redundancy.
We thus discuss some peripherally related works.

The most relevant related work is that of Fan et al.~\cite{diskreduce2}, who
propose to distribute the task of erasure coding using the Hadoop
infrastructure, as MapReduce tasks. Any individual object is however encoded
at a single node, and hence the parallelism achieved in their approach is
only at the granularity of individual data objects. We note from our
experiments that distributing the individual encoding tasks provide further
performance benefits.

Decentralized erasure coding has also been explored in the context of sensor
networks~\cite{decentralizedEC}. However, in such a setting, the (disjoint)
data generated by $k$ sensors is jointly stored over $n>k$ storage sensors
based on erasure coding redundancy. This is achieved using network coding
techniques, and is relatively straight forward to achieve, since random
linear combinations of the already distributed data needs to be stored over
the additional nodes. Such a technique is inapplicable for the problem
considered in this paper.

Li et al.~\cite{Li} also used a similar pipelining based encoding strategy
over a tree-structured topology to reduce the traffic required to repair lost
redundancy. Redundancy replenishment is a very important and vigorously
researched topic~\cite{Li,OD,kermarrec}, however, as noted previously, it is
an unrelated problem.

\section{Conclusions}
\label{s:concl}

In this paper we introduced a novel pipelined erasure coding strategy to
speedup the archival of data in distributed storage systems. We also
presented RapidRAID, an explicit family of erasure codes that realizes the
idea of pipelined erasure coding without compromising either data reliability
or storage overheads. In particular, we showed that for equivalent storage
overhead, RapidRAID codes can achieve a fault tolerance similar to that of
existing erasure codes, and higher than replicated systems. Finally, we
presented a real implementation of RapidRAID codes, and experiments with real
system benchmarks demonstrate the efficacy of our proposed solution. For
coding a single object, our approach achieved up to 90\% reduction in time,
while even when multiple objects are encoded, our approach is up to 20\%
faster than distributing classical erasure coding tasks for different
objects. The benefits of RapidRAID codes are visible even when part of the
network is congested, where RapidRAID codes enable shorter coding times and
have a better scalability as compared to existing erasure codes when the
network congestion increases. The current implementation source code is
available at\linebreak {\em https://github.com/llpamies/ClusterDFS}.

The design of pipelined erasure coding based RapidRAID codes is an important
step towards more efficient mechanisms to archive ``big-data'' in distribute
storage systems. As part of our future research, we aim to explore the
performance of RapidRAID codes under different choice of code parameters $k$
and $n$. It is specially challenging for large values of $n$ where numerical
evaluation of the fault tolerance becomes intractable. We also aim to explore
how RapidRAID codes can be generalized to exploit the existence of more than
two replicas, and particularly for the special case of three replicas, which
is the de facto redundancy scheme used in most production systems.



\end{document}